\documentclass[12pt]{article}
\usepackage{amssymb}
\usepackage{epsfig}

\newcommand{\bq}{\begin{equation}}
\newcommand{\eq}{\end{equation}}
\newcommand{\bqa}{\begin{eqnarray}}
\newcommand{\eqa}{\end{eqnarray}}
\newcommand{\ben}{\begin{enumerate}}
\newcommand{\een}{\end{enumerate}}
\newcommand{\bc}{\begin{center}}
\newcommand{\ec}{\end{center}}

\def\lsim{\lesssim}

\def\pr#1#2#3{ Phys. Rev. ${\bf{#1}}$ (#2) #3}
\def\prl#1#2#3{ Phys. Rev. Lett. ${\bf{#1}}$ (#2) #3}
\def\pl#1#2#3{ Phys. Lett. ${\bf{#1}}$ (#2) #3}

\def\np#1#2#3{ Nucl. Phys. ${\bf{#1}}$ (#2) #3}
\def\zp#1#2#3{ Z. f. Phys. ${\bf{#1}}$ (#2) #3}

\def\etal{{\it et.al.\/}}

\global\nulldelimiterspace = 0pt

\def\O{ {\cal O }}

\begin{document}
\thispagestyle{empty}
\begin {flushleft}

 PM/97-19\\
May 1997\\
\end{flushleft}

\vspace*{2cm}

\hspace*{-0.5cm}
\begin{center}
{\Large {\bf  Residual New Physics Effects in the Heavy Quark
Sector,
Tests at LEP2 and Higher Energies}}
\footnote{{\bf Contribution to the meeting of the european network "Tests of the
Electroweak Symmetry Breaking",
Ouranoupolis, Greece, May 1997.} \\
Partially 
supported by the EC contract CHRX-CT94-0579.}\hspace{2.2cm}\null 

\hspace*{-0.5cm}
\vspace{1.cm} \\{\bf \large F.M. Renard}\hspace{2.2cm}\null \\
 \vspace{0.2cm}  Physique
Math\'{e}matique et Th\'{e}orique, UPRES-A 5032\hspace{2.2cm}\null\\
Universit\'{e} Montpellier
II,  F-34095 Montpellier Cedex 5.\hspace{2.2cm}\null\\[1cm]

\vspace*{0.1cm}
\end{center}
{\bf Summary of works done in collaboration with G.J. Gounaris, 
M. Kuroda,\\
J. Layssac, D.T. Papadamou and C. Verzagnassi.}\\

\vspace*{2cm}
\begin{center}
{\bf Abstract}\hspace{2.2cm}\null
\end{center}
%
%
\begin{minipage}[b]{16cm}
We discuss the sensitivity of the processes $e^+e^-\to t\bar t$ 
and $e^+e^-\to b\bar b$ to special sets of operators describing 
residual New Physics effects. Experimental data in this sector
together with those expected in the bosonic sector should allow to
constrain possible New Physics schemes with effective scales in
the 10 TeV range.
\end{minipage}

\setcounter{footnote}{0} 
\clearpage
\newpage 
  
\hoffset=-1.46truecm
\voffset=-2.8truecm
\textwidth 16cm
\textheight 22cm
\setlength{\topmargin}{1.5cm}


We assume that a certain dynamics generically called New Physics
(NP) exists beyond the Standard Model (SM) and is responsible for many
of its unexplained features like the mass generation mechanism, the
structure of the scalar sector, the mass spectrum of the leptons, the
quarks and the gauge bosons. This new dynamics involves heavy degrees
of freedom caracterized by an effective scale expected to be
much higher than the electroweak scale, i.e. $\Lambda>>M_W$. It is then
natural to expect that the same dynamics which leads to the mass of the
usual particles also generates some contributions to vertices among
these particles. This is what is called residual NP effects. 
This picture suggests that these effects most probably affect particles 
especially concerned by the
mass generation mechanism like $W_L$, $Z_L$ or heavy quarks which may
then act as windows to NP. In addition through gauge interactions 
the other (light)
particles may also get affected.\par
Recent hints for anomalous effects have appeared in $Z\to b\bar b$,
i.e. $\Gamma(Z\to b\bar b)$ is 1.8 $\sigma$ too high and $A_b$ is
2$\sigma$ too low as compared to SM predictions, \cite{databb}. 
There also exist
unexplained features in the high $p_T$ jets produced at 
Tevatron \cite{dijet} and in high
$q^2$ events at HERA \cite{HERA}.\par
To describe these residual effects in a model-independent way we
shall use the effective lagrangian method.
For $M_W\lsim E << \Lambda$ by integrating out
all NP degrees of freedom one obtains an effective Lagrangian
describing the residual effects among usual particles. This Lagrangian
has the following properties. It satisfies $SU(2)\times U(1)$ 
gauge invariance broken at the electroweak scale. The higgs particle
may be light (its mass corresponding to the electroweak scale) or heavy;
this is an open option. The lowest dimensions should be dominant, i.e.
$d=6$. In this report we will only (for simplicity) consider 
CP-conserving effects. With these assumptions one is able to draw
a list of operators \cite{Buch}, 
each of them, $\O_i$, being
associated to a certain dimensionful ($d-4=2$) coupling, $g_i$. 
These couplings are the free
parameters of the description but they can be related to the NP scale by
unitarity arguments \cite{unit}. As a $d=6$ operator 
generates amplitudes
growing with the energy, at some point a saturation of unitarity
occurs. Heavy NP degrees of freedom should appear to cure it (creation
of new particles, resonances,...). It is then natural to identify this
energy value with the effective NP scale.\par
The general list of $\O_i$ involves various classes of bosonic and
fermionic operators with specific features. 
Bosonic operators have been separated
into "non-blind", "blind" and even "super-blind" sets, depending on
their appearence in Z-peak observables at tree or loop levels,
\cite{DeR}, \cite{Hag}, \cite{GGbos}.
The same concept has been extended to the list of heavy quark
operators. In addition
another feature has been added, the presence or not of a 
$t_R$ field (especially involved in mass generation),
\cite{Zbbtop,Kur}. 
Light fermion contributions and left-handed currents should a priori
not be involved.
Equations of motion for fermion fields have been used in order to
eliminate derived operators. We did not do it for those concerning 
the gauge boson fields as
this would bring into the game light fermionic currents\cite{Papa}. 
Operators
involving such
derivative terms will constitute a third class to which we will devote
a separate discussion \cite{bop}. \par
Tests of bosonic operators were discussed in previous reports,
for a review see \cite{LEP2}. Here, we
report on the tests concerning the heavy quark
sector,\cite{Zbbtop,Kur,bop,dj,whisnant,Yuan}.\par
Obviously, the heavy top should be a priviledged place for looking for
these effects, so we start with the process $e^+e^-\to t\bar t$
observable at LC \cite {LC}. Operators of 
Class 1 lead to modifications of the
$\gamma t\bar t$ and $Zt\bar t$ couplings. The general form of such
CP-conserving vertices is given below:
\begin{equation}  \label{eq:dgz}
-i \epsilon_\mu^V J^{\mu}_V = -i e_V \epsilon_\mu^V \bar
u_t(p)[\gamma^{\mu}d^V_1(q^2)+\gamma^{\mu}\gamma^5d^V_2(q^2)
+(p-p^{\prime})^{\mu}d^V_3(q^2)/m_t] v_{\bar t}(p^{\prime}) \ ,
\end{equation}
where $\epsilon_\mu^V$ is the polarization of the vector boson
$V=\gamma, Z$. The
outgoing momenta $(p,~p^{\prime})$ refer to $(t,~\bar t) $ 
respectively and
satisfy $q\equiv p+p^{\prime}$. The normalizations are
determined by $e_{\gamma}\equiv e$ and $e_Z\equiv e/(2s_Wc_W)$,
while $d^V_i$ are in general $q^2$ dependent form factors.
At tree level SM feeds a first set, called set (1), of three couplings: 

\begin{equation} \label{eq:dgzSM}
d^{\gamma, SM0}_1 = {\frac{2}{3}} \ \ , \ \ d^{Z, SM0}_1=
g_{Vt}={\frac{1}{2}}%
-{\frac{4}{3}}s^2_W\ \ , \ \ d^{Z, SM0}_2=-g_{At}=-\,
 {\frac{1}{2}} \ \ . 
\end{equation}
SM at 1-loop and NP lead to
additional $q^2$-dependent contributions to these couplings and to
contributions to the second set, called set (2),
containing 
$d^{\gamma}_2(q^2)$, $d^{\gamma}_3(q^2)$ and $d^{Z}_3(q^2)$.
Departures from the SM (tree + 1-loop) are then defined as:
\bq 
\bar d^V_j \equiv d^V_j-d^{V,SM}_j \ \ \ . \ \label{eq:dvbar}
\eq

The
explicit expressions of the contributions to the $\bar d^V_j$ 
of each operator $\O_i$ at tree
level and at 1-loop are given in ref.\cite{Kur}. 
We will now precisely determine the accuracy at
which they can be observed or constrained.\par

\underline{Tests in $e^+e^-\to t\bar t$}\par
The reaction $e^+e^-\to t\bar t$ offers a way to determine all six
couplings
by studying the top quark spin density matrix elements. In fact the top
quark properties are reconstructed through the decay chain $t\to Wb$,
$W\to l\nu$ or $W\to q\bar q'$. The 4-dimensional angular distribution
of this final state should allow to measure $\rho^t_{\lambda\lambda'}$
at any $q^2$ and production angle $\theta$ as each  of them
leads to a specific angular dependence \cite{Kur, dj}. 
When $e^-$ longitudinal
polarization is available one deals with 
six independent informations $\rho^{L,R}_{++,--,+-}$
that can be expressed in terms of the six $\gamma t\bar t$, $Z
t\bar t$ couplings. Note that absolute measurements
of density matrix elements (like the production cross section
$\sigma_{t\bar t}$) are affected by the lack of precise knowledge of the
$t\to Wb$ decay width \cite{treview, effic}. 
On the opposite, asymmetries 
like forward-backward asymmetries or L-R asymmetries 
(ratios of cross
sections) are free
of this ambiguity. In order to reach pure informations on the
$\gamma t\bar t$, $Z t\bar t$ couplings 
we have tried to rely as much as 
possible on such observables.\par
In the general 6-parameter case this is not fully possible.
For the first set of $\bar d_j$ couplings, asymmetries only depend on
two combinations of $\bar d_j$. To get a third one, informations
sensitive to the absolute normalization, like
$\sigma_{t\bar t}$, are required but they 
will depend on the top quark decay
width. For the second set of couplings we do not have this problem;
asymmetries alone are sensitive to
ratios of the type $\bar d/d^{SM}$ and allow to constrain
the complete set (2). We have made applications to an LC collider 
\cite {LC} with $0.5$,
$1$ and $2$ TeV and luminosity of $20$, $80$ and $320$ $fb^{-1}$
respectively leading to more than $10^{4}$ events. A reconstruction
and detection efficiency of 18 percent \cite{effic} 
has been applied before
computing the statistical accuracy for each
observable. From this we have obtained the accuracy expected in the
determination of the top quark density matrix elements and the
constraints on the NP couplings $\bar d_j$ (ellipsoid in 6-parameter
space). Examples are shown in Fig 1a,b. The complete set of results
is given in
ref.\cite{dj}.\par
The essential features are the following.
For couplings of set (1), as one information is
missing when only asymmetries are used, a band of width $\pm0.02$ 
appears. A
constraint of the order of $\pm0.05$ ($\pm0.02$) is obtained 
when cross section
measurements are added assuming an uncertainty of $20\%$ ($2\%$) 
on the top
decay width. For couplings of set (2), asymmetries alone already
allow to constrain the couplings at the level of $\pm0.02$ in the
unpolarized case and $\pm0.01$ in the polarized case.\par

 We have then studied some constrained cases. From
the list of NP operators, keeping only those contributing at tree level
to the $\gamma t\bar t$ and $Z t\bar t$ couplings,
leaves only four free
parameters associated to  $\O_{t2}$, $\O_{Dt}$, $\O_{tW\Phi}$ and
$\O_{tB\Phi}$.
These contributions affect the $\gamma
t\bar t$ couplings only through $\sigma^{\mu \nu}q_{\nu}$ 
type of vertices,
imposing the relations
\bq
\bar{d}^{\gamma}_1=-2\bar{d}^{\gamma}_3 \ \ , 
\ \ \bar{d}^{\gamma}_2=0\ \ ,
\eq
\noindent
so that the four parameters are $\bar{d}^{\gamma}_3$,
$\bar{d}^{Z}_{1,2,3}$. 

This set is further reduced to only three free parameters if one keeps
only those operators which are generated in the type of dynamical
models studied in ref.\cite {Papa}, namely, $\O_{t2}$,  
$\O_{tW\Phi}$ and $\O_{tB\Phi}$. For these operators, in addition to
the above single photon coupling, only two
$Zt\bar t$ couplings appear, $\sigma^{\mu \nu}q_{\nu}$ and
$\gamma^{\mu}(1+\gamma^5)$, so that:
\bq
\bar{d}^{Z}_1=-2\bar{d}^{Z}_3  \ \ ,
\eq
\noindent
and the three parameters left are $\bar{d}^{\gamma}_3$,
$\bar{d}^{Z}_{2,3}$.

We have also made an illustration with a two-parameter model 
as suggested by
the chiral description which only involves anomalous left-handed and
right-handed $Z t\bar t$
couplings called $\kappa^{NC}_{L,R}$ in ref.\cite{tchiral}.\par
In all these constrained cases the typical accuracy 
for both sets of couplings using only asymmetries
is of the order of $\pm0.02$ without polarization and 
it improves down to $\pm0.01$ when
polarization is available.
Finally we have treated all the operator listed in Class 1, taking them
one by one. Results can be found in Table 7 of ref.\cite{dj}. 
Essentially
there are two levels of constraints. For the four operators 
contributing at tree level, $\O_{t2}$, $\O_{Dt}$, $\O_{tW\Phi}$ and
$\O_{tB\Phi}$, the constraints reach effective scales lying in the 5 to
50 TeV range and are much stronger than the indirect ones already set
by Z-peak results. On the contrary for the operators $\O_{qt}$,
$\O^{(8)}_{qt}$,$\O_{tb}$, that contribute at loop level, the
constraints are expected not to be substancially improved. The scales
lie in the 1-5 TeV range. However two more operators,
that do not substancially contribute at Z-peak, $\O_{tt}$,
$\O_{tG\Phi}$, although contributing at loop level, will get
interesting constraints in the 5-10 TeV range.  \par

\underline{Tests in $e^+e^-\to b\bar b$}\par
We now discuss the $e^+e^-\to b\bar b$ channel. We want to show that
it can give several interesting informations on particular
operators. A priori this could seem unlikely because of the strong
Z-peak constraints in the $b\bar b$ sector. In fact these constraints
come from direct NP effects on $\gamma b\bar b$ and $Zb\bar b$ couplings
ref.\cite{Bing-Lin}; they are induced by operators of Class 2
contributing at tree level. The 0.5 percent accuracy of
$\Gamma(Z\to b\bar b)$ already pushes the NP scale for 
these operators in the
10 TeV range and this cannot be improved by measurements in the
$e^+e^-\to b\bar b$ channel at higher energies. Operators of Class 1
get also constrained through virtual (1-loop) contributions enhanced by
$m^2_t/M^2_W$ factors. This enhancement, well-known in the SM case,
occurs for certain NP operators, see ref.\cite{Zbbtop}. 
In some cases, as we have seen above with $\O_{qt}$,
$\O^{(8)}_{qt}$,$\O_{tb}$, Z peak measurements give 
better constraints than $e^+e^-\to t\bar
t$ at high energies. \par
However, we have found that 
$e^+e^-\to b\bar b$ beyond Z-peak can improve the constraints
for a certain set of "non-blind" operators, i.e. those containing
derivatives of gauge fields (Class 3). For examples a strong
$q^2$ dependence appear in the modification of the gauge boson
propagators due to  $\O_{DW}$, $\O_{DB}$
 ref.\cite{Hag1,clean},
and in the $\gamma b\bar b$, $Zb\bar b$ couplings due to $\O_{qW}$, 
$\O_{qB}$, $\O_{bB}$.
Their effects in $e^+e^-\to b\bar b$ can also be easily computed
through the use of equations of motion. For example in
\bq
\overline{\O}_{DW}  = 2 ~ (D_{\mu} \overrightarrow W^{\mu
\rho}) (D^{\nu} \overrightarrow W_{\nu \rho})  \ \ \
  , \ \  \label{listDW}  \\[0.1cm]
\eq
\noindent
one uses

\bq
D_\mu \overrightarrow W^{\mu\nu}  =  g 
\overrightarrow J^\nu_{(2)}~ -~ i~\frac{g}{2} ~[D^\nu \Phi^\dagger
\overrightarrow \tau \Phi ~ -~ \Phi^\dagger
\overrightarrow \tau D^\nu \Phi ]\ \ , \label{listDmuW}\\[0.1cm] 
\eq
\noindent
$J^\nu_{(2)}$ being the $SU(2)$ current; and similarly for the other
operators, see ref.\cite{Papa}.\\  
They lead to an effective
four-fermion $(e^+e^-)(b\bar b)$ contact interaction. 
This is the origin of the
enhancement of their contribution as compared to the Z tail
contribution which decreases like $1/q^2$. Moreover this property
allows to disentangle their contribution from the ones of all other
operators by using the so-called "Z-peak subtraction method".
Ref.\cite{Zsub}.
It consists in using as inputs $\Gamma(Z\to f \bar f)$, $A_f$
instead of $G_{\mu}$, $s^2_W$. This procedure automatically takes care
of any NP effect at Z peak and leaves only room for those ones which
can survive in the difference of structure functions of the type
\bq  F_i(q^2)-F_i(M^2_Z)
\eq
which are combinations of self-energies, vertices, box and
NP contributions.
 In this way we get contributions from operators of
Class 3, $\overline{\O}_{DB}$, $\overline{\O}_{DW}$, $\O_{qW}$, 
$\O_{qB}$, $\O_{bB}$.
At Z-peak they are mixed with all other operators. Beyond Z-peak 
all other oerators do not contribute.  $\overline{\O}_{DB}$ 
and $\overline{\O}_{DW}$ get already constrained from $e^+e^-\to
l^+l^-$ and $e^+e^-\to q\bar q$. $\O_{qW}$, 
$\O_{qB}$, $\O_{bB}$ can then be studied with
the four observables $\sigma_{b}$,
$A^{b}_{FB}$, $A^b_{LR}$, $A^{pol(b)}_{FB}$ . 
At LEP2 polarization is not available and
we only get two constraints for three operators. This is the origin of
the band appearing in Fig.2 (illustrations for other couples of
operators can be found in ref.\cite{bop}).  
At LC with polarization the system
can be completely constrained leading to sensitivity limits 
on effective scales in the range 30-50 TeV, indeed
a rather high level.\par
In conclusion, we summarize the panorama of the constraints 
already obtained
or expected from future experiments for the whole set of
$dim=6$-operators. We give below the range of NP scales (in TeV) up
to which values experiments can be sensitive.\par
\newpage

 For the bosonic operators :\\

\noindent
 ~~~~ ~~~   non-blind  ~~~~~~~~~~~~~~~~~~~~   17~ (LEP2) 
~~~~~~~~~~~~~   50 ~(LC)\\

~~~~     blind (TGC)~~~~~~~~~~~~~~~~   1.5 ~(LEP2) 
~~~~~~~~~~~~~ 20 ~(LC)\\

~~~~     superblind(Higgs) ~~~~~~~~~~~ 7 ~(LEP2)  
~~~~~~~~~~~~~  20 ~(LC), 
70 (LC$\gamma\gamma$)\\

 For heavy quark operators:\\

\noindent
~~~~~  ~~~      Class 1 ~~~~~~~~~~~      5 ~(LEP1)
~~~~~~~~~     10-50 ~(LC)\\

~~~~~      Class 2 ~~~~~~~~~~        10 ~(LEP1)
~~~~~~~~~~     10 ~(LC)\\

~~~~~      Class 3  ~~~~~~~~~~~       8 ~(LEP2)
~~~~~~~~~~      30-50 ~(LC)\\

\noindent
Work is in progress for the case of the pure gluonic operators
$\O_G$, $\O_{DG}$, $\O_{GG}$.\par
As one can see, limits obtained or expected often lie 
in the 10 TeV range. This is
a domain which covers several types of models beyond SM (Technicolour,
extended gauges,...). Our hope is that experiments 
in various sectors will
reveal certain correlations which could
select a few operators of our list and give  hints for the structure
of new physics.\par

\newpage
\def\x{$\bar d^\gamma_1$}
\def\y{$\bar d^\gamma_2$}
\def\z{$\bar d^\gamma_3$}
\def\u{$\bar d^Z_1$}
\def\v{$\bar d^Z_2$}
\def\w{$\bar d^Z_3 $}

\vspace*{1.cm}
\hspace{3cm} \u
\vspace*{-3cm}
\[ \hspace*{1cm}
\epsfig{file=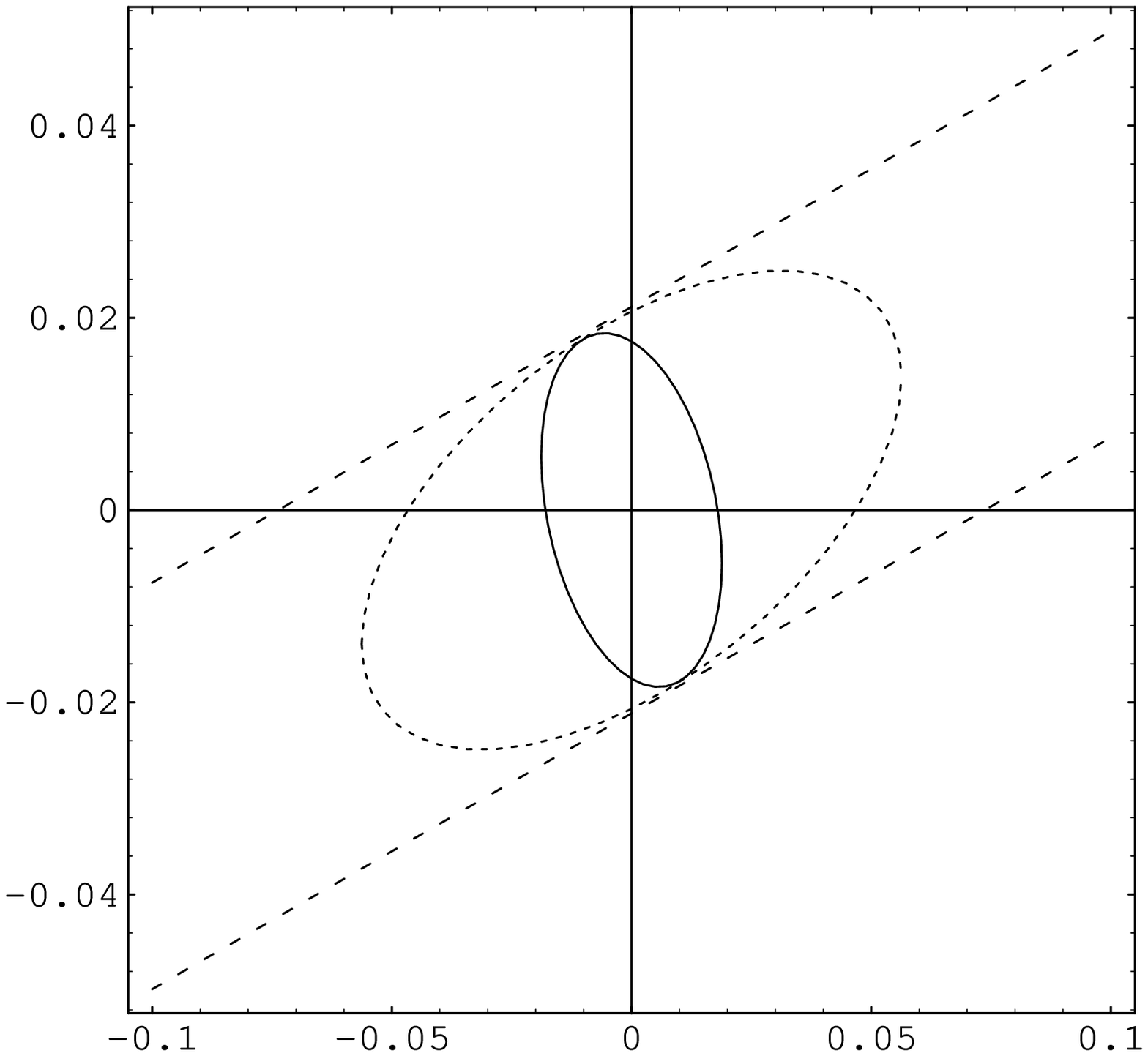,height=12cm}\]\\

\vspace*{-4.cm}
\hspace{12.5cm} \x

\vspace*{0.2cm}
\hspace*{8cm} (a)\\

\vspace*{0.2cm}
\hspace{3cm} \w


\vspace*{-2.7cm}
\[\hspace*{1cm}
\epsfig{file=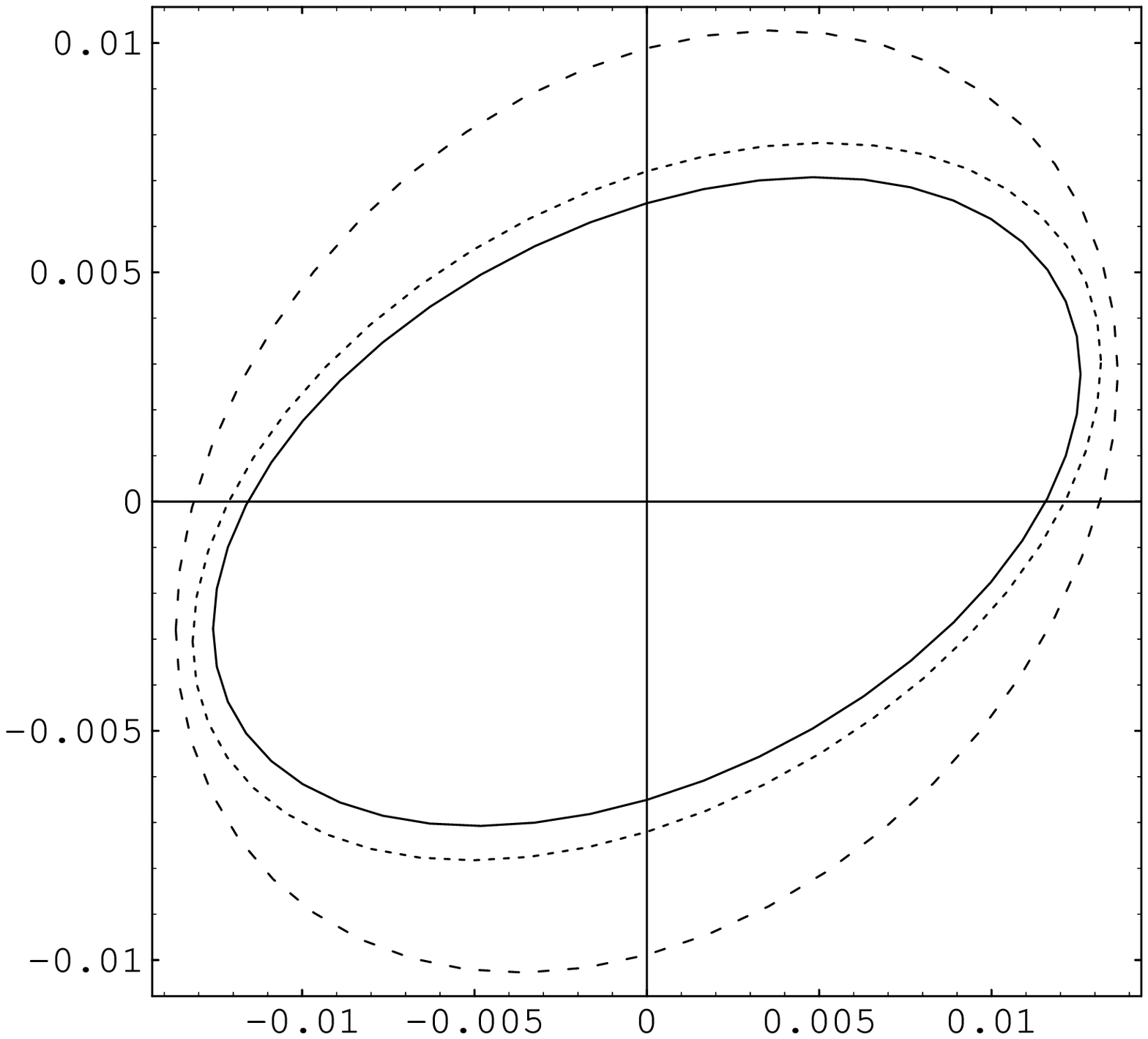,height=12cm}\]\\

\vspace*{-4cm}
\hspace*{12.5cm} \y\\

\hspace*{8cm} (b)

\vspace*{0.5cm}
\begin{center} 
{\bf Fig.1}\hspace{0.5cm} Observability limits in the 6-parameter case;
(a) among couplings of set (1),\\ 
(b) among couplings of set (2);
from asymmetries alone (- - - -), from asymmetries
and integrated observables with a normalization uncertainty of
2\% (........), 20\% (----------).
\end{center} 
\newpage

\vspace*{-3cm}
\hspace*{-3cm}
\epsfig{file=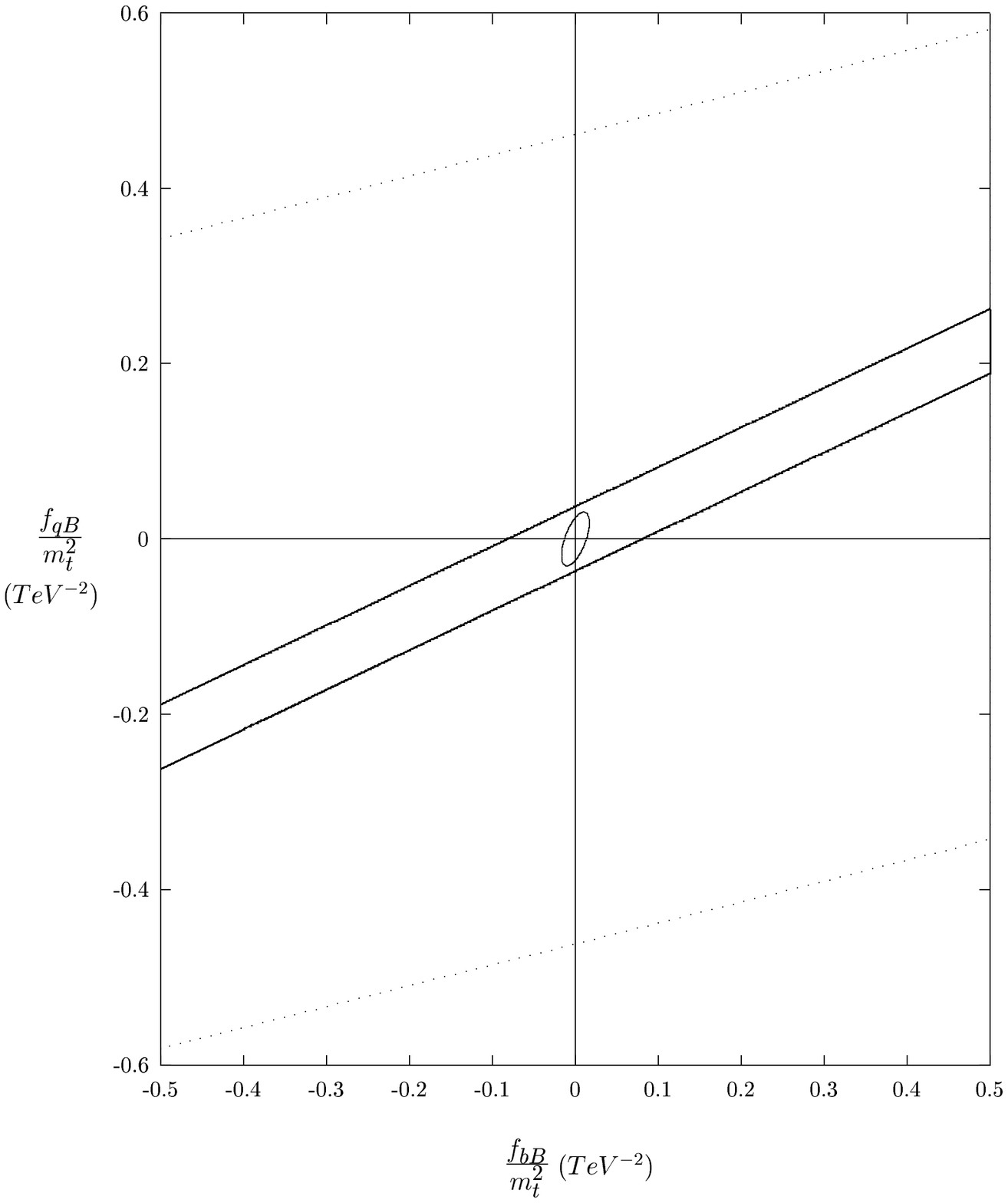,height=27cm}

\vspace*{-5cm}
{\bf Fig.2}\hspace{0.5cm} Constraints from $e^+e^-\to b\bar b$ 
observables in the
3-free parameter case, projected on the ($f_{qB}$, $f_{bB}$) plane; 
at LEP2 (without polarization) ({\it dotted}), ~at 
NLC (without polarization) ({\it solid}),
at NLC (with polarization) ({\it ellipse}).


\begin{thebibliography}{99}

%
\bibitem{databb}
See the talks of
A. Blondel and M.W. Gr\"unewald at the Warsaw conference on High 
Energy Physics (1996).

\bibitem{dijet} CDF, F.Abe et al, \prl{77}{1996}{438}; 
\pr{D55}{1997}{5263}.


\bibitem{HERA} H1 Coll., C. Adloff et al, \zp{C74}{1977}{191};
ZEUS Coll, J. Breitweg et al, \zp{C74}{1997}{207}.
 

\bibitem{Buch} W. Buchm\"{u}ller and D. Wyler,
\np{B268}{1986}{621}; C.J.C. Burgess and H.J. Schnitzer,
\np{B228}{1983}{454}; C.N. Leung, S.T. Love and S. Rao
\zp{C31}{1986}{433}. C. Arzt, M.B. Einhorn and J. Wudka,
\np{B433}{1995}{41}. 
%
 \bibitem{unit} G.J. Gounaris, J. Layssac and F.M. Renard,
\pl{B332}{1994}{146}; G.J. Gounaris, J. Layssac, J.E. Paschalis
and F.M. Renard, \zp{C66}{1995}{619}; G.J. Gounaris, F.M. Renard
and G. Tsirigoti, \pl{B350}{1995}{212};
G.J. Gounaris, F.M.Renard and N.D.Vlachos Nucl.Phys. B459(1996)51;
 M. Hosch, K. Whisnant and B-L. Young,
AMES-HET-96-04 (1996).
%
\bibitem{DeR}
 A. De R\'{u}jula \etal~, \np{B384}{1992}{3}.
%
\bibitem{Hag}
K.Hagiwara et al, \pl{B283}{1992}{353};\pr{D48}{1993}{2182}.
K. Hagiwara, S. Ishihara, R. Szalapski
and D. Zeppenfeld, \pl{B283}{1992}{353} and \pr{D48}{1993}{2182}.

%
 \bibitem{GGbos} G.J. Gounaris, F.M. Renard
and G. Tsirigoti,\pl{B338}{1994}{51}; \pl{B350}{1995}{212}.
%
%
\bibitem{Zbbtop}
 G. Gounaris, F.M.Renard and C.Verzegnassi, 
\pr{D52}{1995}{451}.
%
\bibitem{Kur}
  G.J. Gounaris, M. Kuroda and F.M.Renard, PM/96-22, THES-TP
96/06,  Phys. Rev. $\mathbf{D54}$ (1996) 6861.
%
\bibitem{Papa}
 G.J. Gounaris, D. Papadamou and F.M. Renard, PM/96-28,
THES-TP 96/09, hep-ph/9609437, to appear in Z.f. Physik C.
G.J. Gounaris, D.T. Papadamou and F.M.
Renard, PM/96-31, THES-TP 96/10, hep-ph/9611224, (unpublished). 
%
\bibitem{bop} G.J. Gounaris, D.T. Papadamou and F.M. Renard, THES-TP
97/02, PM/97-02, hep-ph/9703281.

%
\bibitem{LEP2}   Physics at LEP2, Proceedings of the Workshop-Geneva,
Switzerland (1996), CERN 96-01, 
G. Altarelli, T. Sjostrand and F. Zwirner eds.

%
\bibitem{dj}  G.J. Gounaris, D. Papadamou and F.M. Renard, THES-TP
97/02, PM/97-02, hep-ph/9703281.
%
\bibitem{whisnant}
K. Whisnant, J.M. Yang, B.-L. Young and X. Zhang, AMES-HET-97-1,
hep-ph/9702305; A. Data and X. Zhang, hep-ph/9611247.
%
\bibitem{Yuan} C.-P. Yuan, Lectures at the VI Mexican School of 
Particles and Fields, Villahermosa, Mexico 1994, MSUHEP-50228.
%

%
\bibitem{LC} $e^+e^-$ Collisions at 500 GeV: The Physics Potential,
Proceedings of the Workshop-Munich, Annecy, Hamburg, DESY 92-123A (1992),
92-123B (1992), 93-123C (1993), P.M. Zerwas ed.
%
\bibitem{treview} For a review, see e.g., J.H. K\"uhn, TTP96-18 (1996);
F. Larios, E. Malkawi and C.-P. Yuan, MSUHEP-60922, hep-ph/9609482;
D.O. Carlson and C.-P. Yuan, MSUHEP-50823, 1995; C.-P.
Yuan hep-ph/9604434.
%
\bibitem{effic} R. Frey, talk given at Conf. on Phys. and Exp.
with Linear Colliders, Morioka-Appi, Japan (1995), OREXP 96-04. 

%
\bibitem{tchiral} R.D. Peccei and X. Zhang, \np{B337}{1990}{269};
R.D. Peccei, S. Peris and X. Zhang, Nucl. \np{B349}{1991}{305}; 
E. Malkawi and C.-P. Yuan, \pr{D52}{1995}{472};
F. Larios and C.-P. Yuan, hep-ph/9606397.
%
\bibitem{Bing-Lin} X.Zhang and B.L. Young, \pr{D51}{1995}{6584};
A. Datta, K. Whishnant, Bing-Lin Young and X. Zhang,
preprint AMES-HET-97-6.

%
\bibitem{Hag1}  K. Hagiwara, S. Matsumoto, and R. Szalapski,
\pl{\bf B357}{1995}{411}; K. Hagiwara, T. Hatsukano, S. Ishihara
and R. Szalapski, KEK-TH-497(1996).
%
\bibitem{clean}
 A. Blondel, F.M. Renard, L. Trentadue and C.
Verzegnassi,\pr{D54}{1996}{5567};
F.M.~Renard and C.~Verzegnassi, PM/96-27(1996),
to appear in Phys.Rev.D. 



\bibitem{Zsub} F.M.~Renard and C.~Verzegnassi, \pr{D52} 
{1995}{1369}.\pr{D53} 
{1996}{1290}.
%
\bibitem{dynbos}  G.J.Gounaris, F.M.Renard and G.Tsirigoti,  
\pl{B338}{(1994)}{51},\pl{B350}{(1995)}{212}. 
%






\end{thebibliography}
\end{document}